\documentclass[a4paper,11pt]{article}

\usepackage{pos}
\usepackage[capitalise]{cleveref}
\usepackage[utf8]{inputenc}
\usepackage{graphicx}
\usepackage{dsfont}
\usepackage{subcaption}
\usepackage{hyperref}

\graphicspath{{./figures/}}

\newcommand{\ii}{\mathrm{i}}
\newcommand{\obs}{\mathcal{O}}

\title{Complex Langevin simulations with a kernel}

\author*[a]{Michael Mandl}
\author[b]{Erhard Seiler}
\author[a]{D{\'e}nes Sexty}

\affiliation[a]{Institute of Physics, NAWI Graz, University of Graz,\\ Universitätsplatz 5, 8010 Graz, Austria}
\affiliation[b]{Max-Planck-Institut für Physik (Werner-Heisenberg-Institut), \\ Boltzmannstraße 8,
85748 Garching bei München, Germany}

\emailAdd{michael.mandl@uni-graz.at}
\emailAdd{ehs@mpp.mpg.de}
\emailAdd{denes.sexty@uni-graz.at}

\abstract{
	We discuss recent developments regarding the use of kernels in complex Langevin simulations. In particular, we outline how a kernel can be used to solve the problem of wrong convergence in a simple toy model. Since conventional correctness criteria for complex Langevin results are only necessary but not sufficient, the correct convergence of complex Langevin simulations is not always straightforward to assess. Hence, we furthermore discuss a condition for correctness that we have recently derived, which is both necessary and sufficient. Finally, we outline a machine-learning approach for finding suitable kernels in lattice gauge theories and present preliminary results of its application to the heavy-dense limit of QCD.
	}

\FullConference{The 42nd International Symposium on Lattice Field Theory (LATTICE2025)\\
2-8 November 2025\\
Tata Institute of Fundamental Research, Mumbai, India\\}

\begin{document}
\maketitle

\section{Introduction}\label{sec:introduction}
	The complex Langevin method \cite{Par83,Kla83} is a promising approach for solving the sign problem that plagues lattice simulations of various systems of physical interest, such as quantum chromodynamics (QCD) at non-zero baryon density or quantum field theories in real time. It is based on stochastic quantization \cite{PW81}, but requires the complexification of the underlying field space due to the complex actions involved. While the method has produced numerous useful results in various different theories, much is still to be learned about its drawbacks and, most importantly, how to overcome them. In this contribution, we address some of these points. In particular, we outline recent developments dealing with the use of a so-called kernel in the context of the complex Langevin approach that is used to ensure that the method produces correct results. Moreover, we also discuss a novel condition for assessing correctness in the first place.
	
\section{The complex Langevin method}
	Consider a quantum field theory defined by an (in this work: Euclidean) action $S[\Phi]$ depending on some fields $\Phi$. In the conventional approach to lattice field theory, one estimates expectation values of observables $\obs[\Phi]$ via approximating the path integral
\begin{equation}\label{eq:path_integral}
	\langle\obs\rangle = \frac{1}{Z}\int\mathcal{D}\Phi\obs[\Phi]e^{-S[\Phi]}\;, \quad
	Z = \int\mathcal{D}\Phi e^{-S[\Phi]}
\end{equation}
by a sum over a finite number of field configurations distributed according to the path integral weight $e^{-S[\Phi]}$. This importance sampling method, however, breaks down when the action has a non-vanishing imaginary part, as is the case, for instance, in QCD with a non-zero baryon chemical potential, causing a sign problem.

The complex Langevin approach attempts to solve this sign problem via a stochastic evolution of complexified field degrees of freedom in an auxiliary time dimension $\tau$, called the Langevin time. This evolution is governed by the complex Langevin equation, which for a theory of $N$ real scalar fields $\Phi_i^R$ ($i=1,\dots,N$) reads, after complexification $\Phi_i^R\to\Phi_i\equiv\Phi_i^R+\ii\Phi_i^I$,
\begin{equation}\label{eq:cle}
	\frac{\partial\Phi_i(x;\tau)}{\partial\tau} = -H_{ij}H^T_{jk}\frac{\delta S[\Phi]}{\delta\Phi_k(x;\tau)} + H_{ij}\eta_j(\tau)\;.
\end{equation}
Here, $\eta_j$ is a real Gaussian noise vector satisfying $\langle\eta_i(\tau)\rangle=0$ and $\langle\eta_i(\tau)\eta_j(\tau')\rangle=2\delta_{ij}\delta(\tau-\tau')$ for all $i$ and $j$. Moreover, we have introduced a so-called kernel $H_{ij}$. While, in general, this matrix-valued quantity could depend on the fields $\Phi_i$, in which case there would be an additional term in \eqref{eq:cle}, we only consider field-independent kernels here. Note that the summation over indices occurring twice in a product is assumed throughout.

The solution of \eqref{eq:cle} gives rise to a probability density $P(\Phi^R,\Phi^I;\tau)$ in complex $\Phi$-space, whose $\tau$-dependence is given by an associated Fokker--Planck equation. Assuming that its equilibrium limit
\begin{equation}\label{eq:equilibrium}
	P(\Phi^R,\Phi^I) \equiv \lim_{\tau\to\infty}P(\Phi^R,\Phi^I;\tau)
\end{equation}
exists, the complex Langevin approach provides estimators for expectation values of observables via the integration of the real density $P(\Phi^R,\Phi^I)$ over the complex domain as
\begin{equation}
	\langle\obs\rangle_\mathrm{CL} \equiv \int\mathcal{D}\Phi^R\mathcal{D}\Phi^I\obs[\Phi^R+\ii\Phi^I]P(\Phi^R,\Phi^I)\;.
\end{equation}
Thus, the complex Langevin evolution solves the sign problem if and only if
\begin{equation}\label{eq:correctness}
	\langle\obs\rangle_\mathrm{CL}=\langle\obs\rangle\;, 
\end{equation}
with $\langle\obs\rangle$ given in \eqref{eq:path_integral}. We mention in passing that the holomorphicity of the observables $\obs[\Phi]$ as well as of the path integral density $e^{-S[\Phi]}$ is a necessary prerequisite for the method.

\section{The wrong convergence problem}
	It is a well known problem of the complex Langevin evolution that \eqref{eq:correctness} does not always hold in practise. The possible reasons for this failure have been investigated on many occasions and can be summarized as follows:
\begin{itemize}
	\item The probability density $P(\Phi^R,\Phi^I)$ may not decay fast enough in complex field space. This causes the appearance of boundary terms at infinity when trying to perform integration by parts, spoiling a formal proof of correctness of the method \cite{ASS10,AJS11}. While such boundary terms can -- in principle -- be measured in a simulation \cite{SSS19,SSS20}, dealing with them is a much harder task.
	\item In the absence of boundary terms, the complex Langevin evolution may suffer from contributions from unwanted so-called integration cycles \cite{GP09,SS19,HMS25}, which are associated with spurious solutions of the theory's Dyson--Schwinger equations \cite{GGG96}.
\end{itemize}

Here, we shall focus on the second problem in particular. For simplicity, let us follow \cite{GP09} and consider a theory with only a single degree of freedom $z$, for which the Dyson--Schwinger equations read
\begin{equation}\label{eq:dses}
	\left\langle Az^n\right\rangle = 0\;, \quad A = \frac{\partial}{\partial z} - \frac{\partial S(z)}{\partial z}\;,
\end{equation}
where $n$ is a non-negative integer. They can be derived from the generating functional
\begin{equation}\label{eq:Z}
	Z(j) = \int_{-\infty}^{\infty} dz e^{-S(z)+jz}\;.
\end{equation}
depending on the source $j$, via the master equation
\begin{equation}\label{eq:master_equation}
	\frac{\partial S(z)}{\partial z}\bigg\vert_{z=\frac{\partial}{\partial j}}Z(j)=jZ(j)\;.
\end{equation}
The important point to note is that the solution \eqref{eq:Z} to \eqref{eq:master_equation} is not unique. In fact, any 
\begin{equation}
	Z_\gamma(j) \equiv \int_\gamma dz e^{-S(z)+jz} \quad 
	\textnormal{with} \quad
	e^{-S(z)+jz}\bigg\vert_{\partial\gamma}=0
\end{equation}
solves the master equation. Such integration contours $\gamma$ are referred to as integration cycles. Thus, the complexification of the underlying field space enlarges the solution space of a theory's Dyson--Schwinger equations, which has profound consequences for complex Langevin expectation values. 

In particular, it was shown for one-dimensional systems in \cite{SS19} that if \eqref{eq:dses} holds in a complex Langevin simulation, the complex Langevin expectation values are linear combinations of expectation values computed along each of the theory's $n_\gamma$ linearly independent integration cycles, i.e.,
\begin{equation}\label{eq:sase}
	\langle z^n\rangle_\mathrm{CL} = \sum_{i=1}^{n_\gamma}a_i\langle z^n\rangle_{\gamma_i}\;, \quad \textnormal{with} \quad
	\langle z^n\rangle_{\gamma_i} \equiv \frac{\partial\ln(Z_\gamma(j))}{\partial j}\bigg\vert_{j=0} \quad
	\textnormal{and} \quad a_i\in\mathbb{C}\;.
\end{equation}
Thus, defining $\gamma_1=\mathbb{R}$ without loss of generality, complex Langevin produces correct (i.e., physical) results if and only if $a_i=\delta_{i1}$. In \cite{HMS25}, it was argued that one may -- at least to a certain extent -- use a kernel to control which integration cycles are being sampled in a simulation. Moreover, the same reference also established first numerical evidence for the validity of \eqref{eq:sase} for more than one degree of freedom. 

The crucial observation is that conventional correctness criteria for complex Langevin do not seem to be sensitive to contributions from unwanted integration cycles: the Dyson--Schwinger equations hold and boundary terms vanish, but the results complex Langevin produces are nonetheless incorrect. We also mention that the correctness criterion based on the decay properties of the complex Langevin drift term \cite{NNS16} also fails in this respect, which we are planning to investigate on a quantitative level in future work. A reliable correctness criterion is thus much needed.

\section{A novel condition for correctness}
	In \cite{MSS25}, we have worked out a family of necessary and sufficient conditions for correctness of complex Langevin simulation results. We demonstrate them here in a simplified manner, using a particular one-dimensional toy model as an example. For a full derivation in general dimensions, a more detailed discussion of the underlying assumptions, as well as applications to a handful of other models, we refer to \cite{MSS25}. The theory we shall consider here is given by the action
\begin{equation}\label{eq:model}
	S(z) = \frac{\lambda}{4}z^4\;, \quad \lambda=e^{\frac{5\ii\pi}{6}}\;.
\end{equation}
It has been investigated before in the context of complex Langevin simulations, in particular regarding the application of a kernel, on multiple occasions \cite{OOS89,MHS24p,MSS25}. 

In order to outline the correctness conditions, let us first define the $p$-norms (with $p\geq1$) of functions $f(z)$:
\begin{equation}
	\Vert f\Vert_p = \left(\int_{\mathcal{M}_r}dz\vert f(z)\vert^p\right)^{1/p} \quad \textnormal{and} \quad \Vert f\Vert_\infty = \sup_{z\in\mathcal{M}_r}\vert f(z)\vert\;,
\end{equation}
where $\mathcal{M}_r$ can be chosen to be any real manifold that contains no zeros of $e^{-S(z)}$. In most applications it will either be the real line $\mathbb{R}$ or some continuous deformation of $\mathbb{R}$ to avoid such zeros. Next, we introduce the functions $w(z)$ and $\rho_r(z)$ as
\begin{equation}
	w(z) = e^{-\frac{\lambda}{8}z^4}\;, \quad \rho_r(z) = \frac{1}{Z}e^{-\frac{\lambda}{8}z^4}\;,
\end{equation}
which gives $\frac{e^{-S(z)}}{Z}=w(z)\rho_r(z)$. Note that this choice of $w(z)$ and $\rho_r(z)$ is by no means unique; see \cite{MSS25} for further details and the precise requirements on $w(z)$ and $\rho_r(z)$. With this, we then define the constants $C^{(p)}$ and the $w$-norms $\Vert f\Vert_w^{(p)}$ as
\begin{equation}
	C^{(p)} \equiv \Vert\rho_r\Vert_p \quad \textnormal{and} \quad
	\Vert f\Vert_w^{(p)} \equiv \Vert w(z)f(z)\Vert_{\frac{p}{p-1}}\;,
\end{equation}
respectively. It was shown in \cite{MSS25} that a necessary and sufficient condition for correctness of complex Langevin results in the model \eqref{eq:model} is that, on the one hand, the Dyson--Schwinger equations of \eqref{eq:model} hold and, on the other hand, the following bounds are obeyed for any choice of $p\geq1$, both for all $f(z)$ in the space of polynomials:
\begin{equation}\label{eq:bounds}
	\left\vert\langle f(z)\rangle_\mathrm{CL}\right\vert \leq C^{(p)}\Vert f\Vert_w^{(p)}\;.
\end{equation} 
Needless to say, it is impossible in practice to verify the validity of \eqref{eq:bounds} for all polynomials, which severely limits the usefulness of the proposed correctness criterion. However, we have found it helpful as a tool for detecting wrong convergence. To be clear, \eqref{eq:bounds} really defines an entire family of conditions, since the choice of $p$, as well as that of the functions $w(z)$ and $\rho_r(z)$ and the real manifold $\mathcal{M}_r$, is essentially arbitrary. This means that if \eqref{eq:bounds} is violated for even a single such choice and any $f(z)$, one concludes that the complex Langevin results are necessarily incorrect. It should be emphasized, however, that what is really meant by ``incorrect results'' is that \eqref{eq:correctness} is violated for at least a single observable. Nonetheless, it is conceivable that even in a scenario in which some bounds in \eqref{eq:bounds} do not hold, certain observables (for instance low powers of $z$) could still come out right. In other words, it is not sufficient to check \eqref{eq:bounds} for a finite set of observables in order to draw conclusions about the correctness of that set.

We have successfully tested this new correctness condition in \cite{MSS25} using $p=1$. Here, however, we shall investigate more general choices $p\geq1$. As in previous works, we parametrize the kernel in \eqref{eq:cle}, which is just a single complex number in the present case, as
\begin{equation}\label{eq:kernel}
	H = e^{-\frac{\ii m\pi}{48}}\;, \quad m=0,1,\dots,47\;,
\end{equation}
and we study results as a function of the kernel parameter $m$. As is known from \cite{OOS89}, complex Langevin produces correct results in the model \eqref{eq:model} for a sizeable range of $m$ close to $m=10$. However, there is another such plateau of equal size around $m=34$, as well as two smaller plateaus near $m=22$ and $m=46$, respectively, on which the Dyson--Schwinger equations are fulfilled but the results are nonetheless incorrect due to contributions from unwanted integration cycles \cite{MHS24p,HMS25}. To the best of our knowledge, the condition \eqref{eq:bounds} is the only correctness criterion that exists to date that could detect this incorrectness without the need of computing exact results.

\begin{figure}[t]
\centering
\begin{subfigure}{.495\linewidth}
	\includegraphics[scale=0.46]{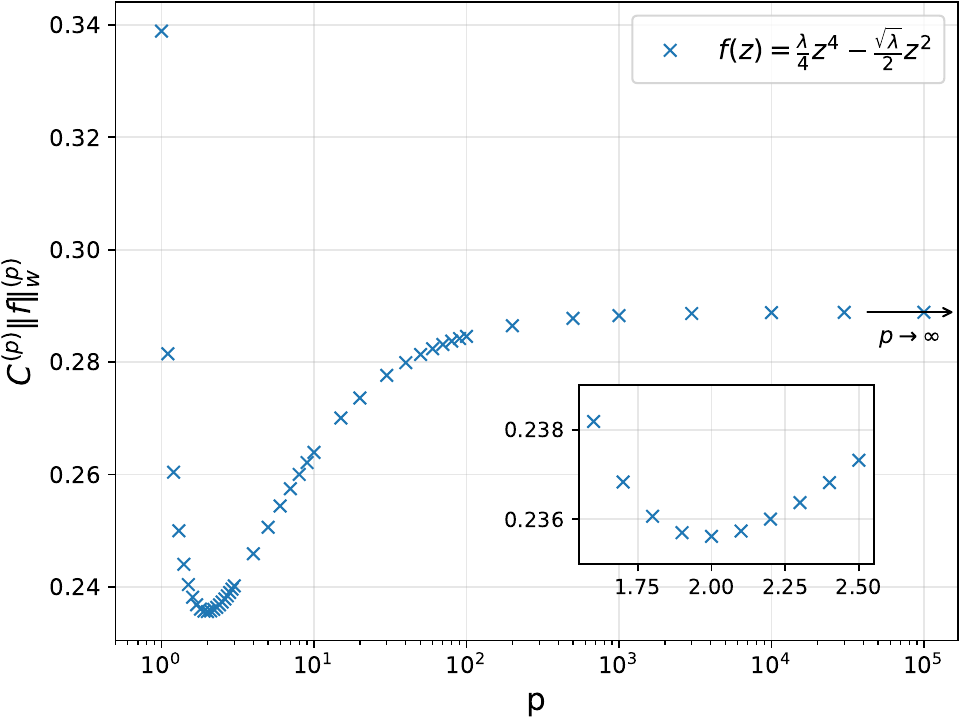}
\end{subfigure}
\begin{subfigure}{.495\linewidth}
	\includegraphics[scale=0.46]{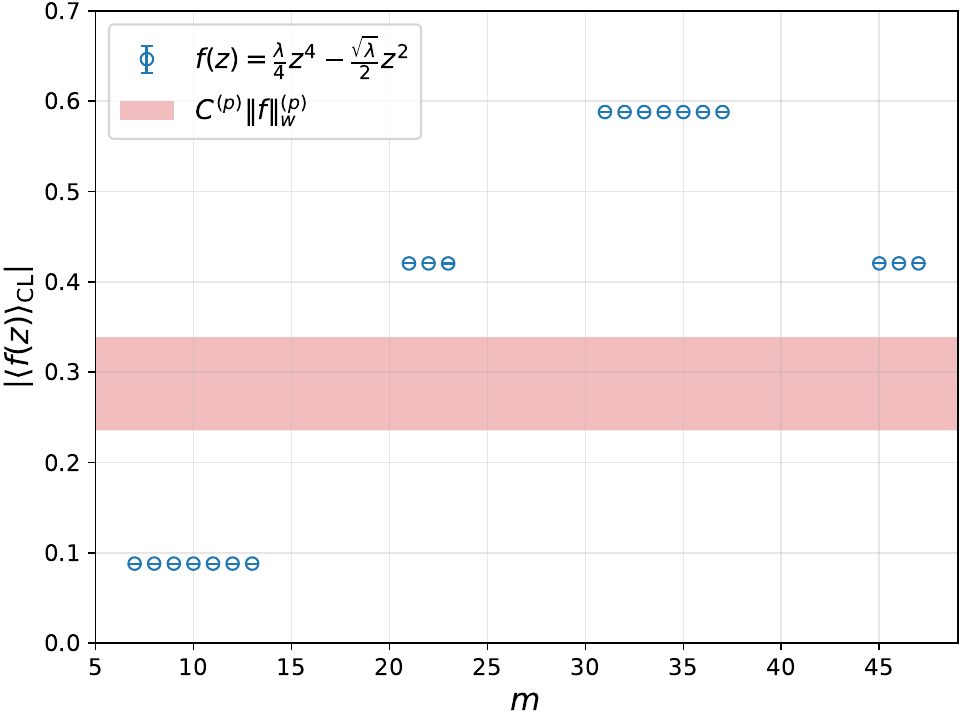}
\end{subfigure}
\caption{Validity of the bounds \eqref{eq:bounds} with $f(z)$ given by \eqref{eq:control_observable}. \emph{(left)}: Dependence of $C^{(p)}\Vert f\Vert_w^{(p)}$ on $p$ on a logarithmic scale. The arrow indicates the limit $p\to\infty$, while the inset shows the region around $p=2$ on a linear scale. \emph{(right)}: $\vert\langle f(z)\rangle_\mathrm{CL}\vert$ as a function of the kernel parameter $m$ in \eqref{eq:kernel}. The red band indicates the range of bounds obtained by varying $p$. For values of $m$ in between the four plateaus, there are non-vanishing boundary terms and the Dyson--Schwinger equations are not fulfilled \cite{MSS25}, such that we refrain from showing any data there.}
\label{fig:criterion}
\end{figure}

To begin with, we plot in \cref{fig:criterion} (left) the dependence of the bound $C^{(p)}\Vert f\Vert_w^{(p)}$ on $p$ for the particular choice
\begin{equation}\label{eq:control_observable}
	f(z) = \frac{\lambda}{4}z^4-\frac{\sqrt{\lambda}}{2}z^2\;.	
\end{equation} 
Curiously, the bound attains a minimum around $p=2$ before converging to its $p\to\infty$ limit monotonically. While each value of $p$ gives rise to an independent condition for correctness, in practise one should of course try to minimize the bounds, in order to have as strong a condition as possible. In the present case, the optimal value is thus $p\approx2$, while the choice $p=1$ in \cite{MSS25} is actually the largest such bound. Note, however, that the bounds for different control observables $f(z)$, as well as for different choices of $w(z)$, $\rho_r(z)$, and $\mathcal{M}_r$ will generally assume minima at different values of $p$. In the following we shall nonetheless consider the full range of bounds \eqref{eq:bounds} obtained by varying $p$. In \cref{fig:criterion} (right), we show the dependence of $\vert\langle f(z)\rangle_\mathrm{CL}\vert$ on $m$ and compare with \eqref{eq:bounds}. One finds that the bounds are violated on all plateaus except the one around $m=10$, on which the results are known to be correct. Thus, our criterion is indeed capable of ruling out all incorrect results. Interestingly, this observation holds independently of the choice of $p$. We emphasize once again that these findings on their own cannot prove the correctness around $m=10$.

\section{Machine-learning kernels in lattice QCD}
	One of the conclusions that can be drawn from the previous section is that the use of a kernel in a complex Langevin simulation can have a large impact on the $\tau$-evolution and the associated equilibrium distribution $P$ in \eqref{eq:equilibrium}. Most importantly, it can potentially give rise to correct results in cases where the method would fail without a kernel. The important question for practical applications is then how to choose appropriate kernels in realistic theories of physical interest, QCD at finite baryon density being just one example. One way to find such stabilizing kernels, which has proven successful in the past \cite{ALR23,ARS24}, is to employ machine-learning techniques. In the following, we briefly summarize the key idea and present a few preliminary results within the heavy-dense limit of QCD. A more detailed analysis is deferred to future work.

In the case of lattice gauge theories, the gauge degrees of freedom are represented by the group-valued link variables $U_\mu(x)$ connecting the lattice site $x$ with its nearest neighbor in $\mu$-direction. For QCD, the relevant gauge group is $\mathrm{SU}(3)$. The complexification required for the application of the complex Langevin method then takes the gauge links outside of this compact domain and into $\mathrm{SL}(3,\mathbb{C})$. More precisely, for a lattice link $X=(x,\mu)$, a naive discretization of the complex Langevin evolution equation for $U_X$ reads
\begin{equation}\label{eq:link_update}
	U_X \to U_X' = \exp\left[\ii\lambda_a\left(-\varepsilon H_{Xa,Yb}H^T_{Yb,Zc}D_{Zc}S+\sqrt{\varepsilon}H_{Xa,Yb}\eta_{Yb}\right)\right]U_X\;.
\end{equation}
Here, $\lambda_a$ are the Gell-Mann matrices, $\varepsilon$ denotes the discrete Langevin-time step size, $D_{Zc}S[U]$ is the usual left-derivative of the lattice action $S[U]$, $\eta_{Yb}$ is Gaussian noise with the same properties as above and $H_{Xa,Yb}$ denotes the kernel matrix to be determined.

A notion of distance between the $\mathrm{SL}(3,\mathbb{C})$ gauge configurations $U_X$ and the target group manifold $\mathrm{SU}(3)$ is given by the so-called unitarity norm
\begin{equation}\label{eq:unitarity_norm}
	N_U \equiv \frac{1}{V}\left[\sum_{X}\left(U_XU_X^\dagger-\mathds{1}\right)^2\right]\;,
\end{equation}
with $V$ denoting the number of lattice points. A common failure mode of complex Langevin simulations in QCD at non-zero chemical potential is that the unitarity norm becomes too large, indicating the sampling of regions of complexified field space that are irrelevant for the path integral. To counteract the growth of $N_U$, we use it as the loss function in a gradient descent approach, with the kernel elements $H_{Xa,Yb}$ as optimization parameters. As a first step, for simplicity, we choose the kernel such that it is the same for every lattice link, i.e., $H_{Xa,Yb}=\delta_{XY}H_{ab}$, and perform the optimization with respect to the $8\times8$ complex matrix elements $H_{ab} = A_{ab}+\ii B_{ab}$, considering their real and imaginary parts $A_{ab}$ and $B_{ab}$ separately. Concretely, the gradient descent step reads 
\begin{equation}\label{eq:gradient}
	H_{ab} \to H_{ab} -l\left(\frac{\partial N_{U'}}{\partial A_{ab}}+\ii\frac{\partial N_{U'}}{\partial B_{ab}}\right) = H_{ab}-2l\frac{\partial N_{U'}}{\partial H^*_{ab}}\;,
\end{equation}
where $l>0$ is a real parameter called the learning rate that we may choose freely. Notice that we are taking the derivatives of $N_{U'}$, which is \eqref{eq:unitarity_norm} with the updated links \eqref{eq:link_update} inserted.

Our training strategy is as follows: We start a simulation with a trivial kernel $H_{ab}=\delta_{ab}$ and let it thermalize for a Langevin-time duration $\tau_\mathrm{therm}$. We then compute the gradient on the right-hand side of \eqref{eq:gradient} multiple times, namely after every $\tau_\mathrm{grad}$ Langevin time for a duration $\tau_\mathrm{update}$, after which we average over these gradients and update the kernel according to \eqref{eq:gradient} using this average. Before the update, the kernel is normalized to keep its (Frobenius) norm constant. With the updated kernel, we then go back to computing the gradients, but without an additional equilibration.

In the following, we test this approach for the machine-learning of kernels in the context of QCD in the heavy-dense limit \cite{BHK92p}. In this setting, the fermionic determinant $\det(M)$ for Wilson fermions reduces to the simple form 
\begin{equation}
	\det(M) = \prod_\mathbf{x}\det\left(\mathds{1}+he^{\mu N_t}\mathcal{P}_\mathbf{x}\right)^2\det\left(\mathds{1}+he^{-\mu N_t}\mathcal{P}_\mathbf{x}^{-1}\right)^2\;,
\end{equation}
where the product runs over the spatial lattice volume, $h=(2\kappa)^{N_t}$ ($\kappa$ is the usual hopping parameter), $N_t$ denotes the number of lattice points in time direction, and $\mu$ is the chemical potential (in lattice units). Finally, the Polyakov loop and the inverse Polyakov loop are defined via the temporal gauge links $U_0(x)=U_0(t,\mathbf{x})$ as
\begin{equation}
	\mathcal{P}_\mathbf{x} = \prod_{t=0}^{N_t-1}U_0(t,\mathbf{x})\;, \quad 
	\mathcal{P}_\mathbf{x}^{-1} = \prod_{t=N_t-1}^0U_0^{-1}(t,\mathbf{x})\;,
\end{equation}
respectively. In our test scenario, we employ a simple Wilson gauge action and the following set of parameters: $\kappa=0.12$, $N_t=4$, $N_s=8$ (with $N_s$ denoting the number of lattice points in the spatial directions), and $\mu=0.8\;$. Moreover, for the inverse gauge coupling $\beta=6/g^2$, we choose the value $\beta=6.0$. Since we are conducting a proof-of-principle study, we do not perform any sort of scale setting, let alone a continuum or infinite-volume extrapolation for the time being. The hyper-parameters we employ in the training are given by $l=10^3$, $\tau_\mathrm{therm}=10$, $\tau_\mathrm{grad}=0.01$, and $\tau_\mathrm{update}=1$, leaving us with a batch size of $100$. Finally, we use a Langevin step size of $\varepsilon=10^{-5}$. We do not employ any improved solvers \cite{UF85} or adaptive step-size algorithms \cite{AJS10}, but we do perform $16$ gauge-cooling steps \cite{SSS13} after every Langevin update. We would like to stress at this point the possibility that the usage of gauge-cooling updates might be in conflict with the idea of using a homogeneous and isotropic kernel in the way outlined above, since our choice of kernel is not invariant under gauge transformations. The effects of this, as well as potential cures, need to be investigated and we are planning to do so in future work.

\begin{figure}[t]
\centering
\begin{subfigure}{.495\linewidth}
	\includegraphics[scale=0.46]{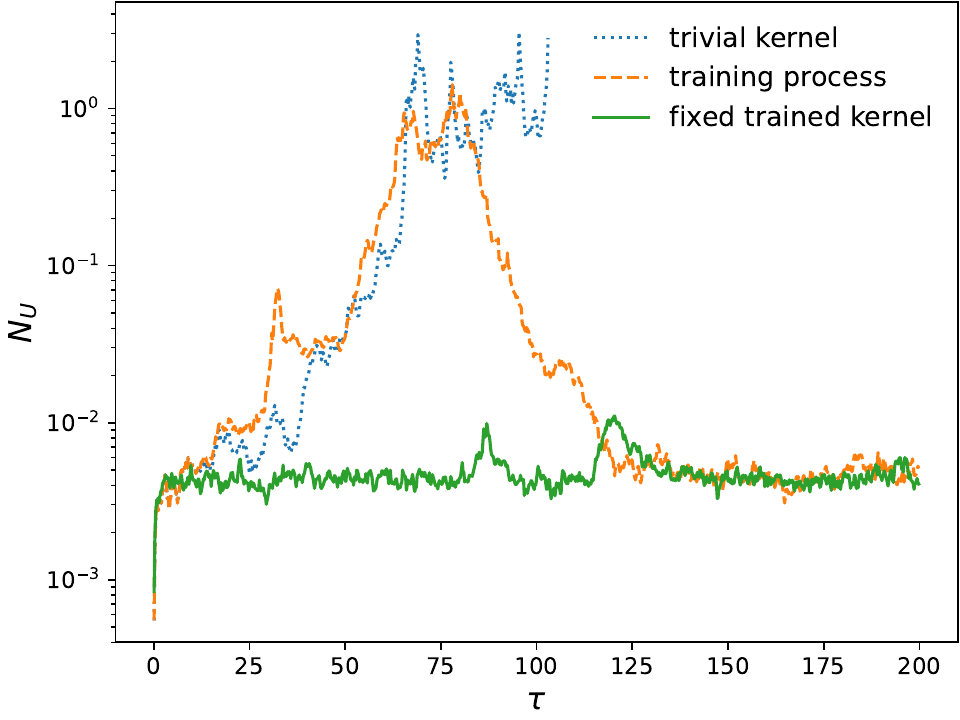}
\end{subfigure}
\begin{subfigure}{.495\linewidth}
	\includegraphics[scale=0.46]{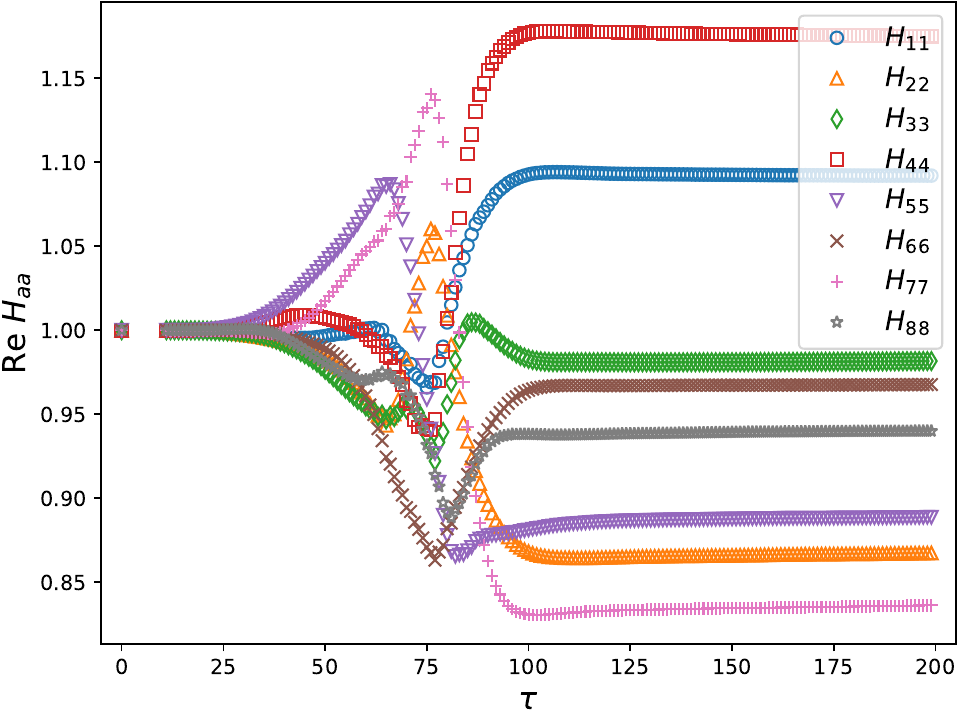}
\end{subfigure}
\caption{Langevin-time evolution of the training process for the complex Langevin kernel; see the main text for the choice of parameters. \emph{(left)}: Unitarity norm \eqref{eq:unitarity_norm} in simulations with a trivial kernel (dotted line), an active training process (dashed line) and a fixed trained kernel (solid line). The initial conditions are the same for each simulation. \emph{(right)}: Real parts of the diagonal kernel elements during the training process.}
\label{fig:training}
\end{figure}

In \cref{fig:training} (left), we show the results of our optimization procedure. In particular, we show the evolution of $N_U$ with the Langevin time in a typical simulation. As can be seen, with a trivial kernel the unitarity norm grows with $\tau$ rather quickly due to the constant step size and the lack of improvement (dotted curve). At a point where it becomes too large to draw meaningful conclusions from the simulation, the latter is terminated.  Once we apply our training procedure (using the same initial configuration and random number seed here for better comparison), however, we observe a significant improvement in the $\tau$-dependence of $N_U$: While initially it grows at about the same rate as with a trivial kernel, it then goes back to small values due to the trained kernel (dashed curve). Most strikingly, the stable behavior observed for large enough Langevin times $\tau\gtrsim130$ indicates that the unitarity norm has reached equilibrium. This is very much desirable, as it extends the range in $\tau$ from which one may reliably draw samples and thus the statistics by a significant amount.

To get an idea about what exactly the kernel learns, we show in \cref{fig:training} (right) the $\tau$-evolution of the (real parts of the) diagonal kernel elements $H_{aa}$. They are initialized as $H_{aa}=1$ and at first vary only slowly with Langevin time. However, at the point where the unitarity norm becomes large, the kernel elements also change faster. Finally, for large enough Langevin times, the $H_{aa}$ show almost no variation as a function of $\tau$ anymore, supporting the picture of equilibration. The real parts of the non-diagonal elements exhibit an analogous behavior, but stay below $\sim0.2$ in magnitude for the entire simulation. A similar statement can be made for the imaginary parts of the kernel elements. All in all, the kernel is still predominantly diagonal after the training process.

In practice, of course, one would rather work with a single fixed kernel to avoid computing gradients for the entire duration of a simulation. To this end, one may use the final kernel matrix $H_{ab}$ that was trained in the above way. In particular, according to \cref{fig:training} (right) that final kernel appears to be stable as a function of $\tau$. The results of a simulation with this fixed trained kernel are shown in \cref{fig:training} (left) as well (solid line). As can be seen, the unitarity norm stays below $\sim10^{-2}$ for the entire simulation, i.e, the simulation trajectory never seems to sample irrelevant regions of the complexified field space. Thus, it appears as if there are no issues related to the use of both a (homogeneous and isotropic) kernel and gauge cooling in the present case. If such a behavior is reproducible for smaller $\beta$, corresponding to lower temperatures, and, in particular, in full QCD, it has the potential to greatly advance the range of applicability of the complex Langevin method. In particular, it could allow the study of previously unexplored regions of the QCD phase diagram, especially when combined with improved solvers and/or adaptive step size algorithms.

\acknowledgments
We are grateful to Enno Carstensen and Ion-Olimpiu Stamatescu for valuable discussions. This research was funded in whole, or in part, by the Austrian Science Fund (FWF) [\href{https://doi.org/10.55776/P36875}{10.55776/P36875}]. The numerical simulations underlying this work were performed on the computing cluster of the University of Graz (GSC).
    
\bibliographystyle{JHEP}
\bibliography{bibliography}

\end{document}